\documentclass[pre,twocolumn,groupedaddress,showpacs,showkeys,amsmath]{revtex4}

\usepackage{graphicx}
\usepackage{dcolumn}
\usepackage{bm}

\begin{document}


\title{The upper critical dimension of the negative-weight percolation problem}

\author{O. Melchert$^1$}
\email{oliver.melchert@uni-oldenburg.de}
\author{L. Apolo$^{1,2}$}
\email{lapolo00@ccny.cuny.edu}
\author{A. K. Hartmann$^1$}
\email{alexander.hartmann@uni-oldenburg.de}
\affiliation{
$^1$ Institut f\"ur Physik, Universit\"at Oldenburg, 26111 Oldenburg, Germany\\
$^2$ City College of the City University of New York, New York, New York 10031, USA
}

\date{\today}


\begin{abstract}
By means of numerical simulations we investigate the geometric properties of loops 
on hypercubic lattice graphs in dimensions $d\!=\!2$ through $7$, where edge weights 
are drawn from a distribution that allows for positive and negative weights. 
We are interested in the appearance of system-spanning loops of total negative 
weight.  The resulting negative-weight percolation (NWP) problem is 
fundamentally different from conventional percolation, as we have seen in 
previous studies of this model for the $2d$ case.
Here, we characterize the transition for hypercubic systems, where the aim of 
the present study is to get a grip on the upper critical dimension $d_u$ of 
the NWP problem.

For the numerical simulations we employ a mapping of the NWP model to a 
combinatorial optimization problem that can be solved exactly by using
sophisticated matching algorithms.
We characterize the loops via observables similar to those in percolation theory 
and perform finite-size scaling analyses, e.g.\ $3d$ hypercubic systems with side 
length up to $L\!=\!56$ sites, in order to estimate the critical properties of the 
NWP phenomenon. We find our numerical results consistent with an upper critical 
dimension $d_u\!=\!6$ for the NWP problem.
\end{abstract} 

\pacs{64.60.ah, 75.40.Mg, 02.60.Pn, 68.35.Rh}
\maketitle

\section{Introduction \label{sect:introduction}}

The statistical properties of lattice-path models on graphs,
equipped with quenched disorder, have experienced much attention during the 
last decades.
They have proven to be useful in order to characterize, e.g.,
linear polymers in disordered/random media \cite{kremer1981,kardar1987,derrida1990,grassberger1993,parshani2009}, 
vortices in  high $T_c$ superconductivity \cite{pfeiffer2002,pfeiffer2003}, 
and 
domain-wall excitations in disordered media such as spin glasses \cite{cieplak1994,melchert2007} and 
the solid-on-solid model \cite{schwarz2009}. 
The precise computation of these paths can often be formulated in terms
of a combinatorial optimization problem and hence might allow for the
application of exact optimization algorithms developed in computer science.

For an analysis of the statistical properties of these lattice-path models, 
geometric observables and scaling concepts similar to those developed in 
percolation theory \cite{stauffer1979,stauffer1994} have been used conveniently. 
In the past decades, a large number of percolation problems in various
contexts have been investigated through numerical simulations.
Among these are problems where the fundamental entities are string-like, 
as for the lattice path models mentioned in the beginning, 
rather than clusters consisting of occupied nearest neighbor sites, 
as in the case of usual random bond percolation.

\begin{figure}[t!]
\centerline{
\includegraphics[width=1.\linewidth]{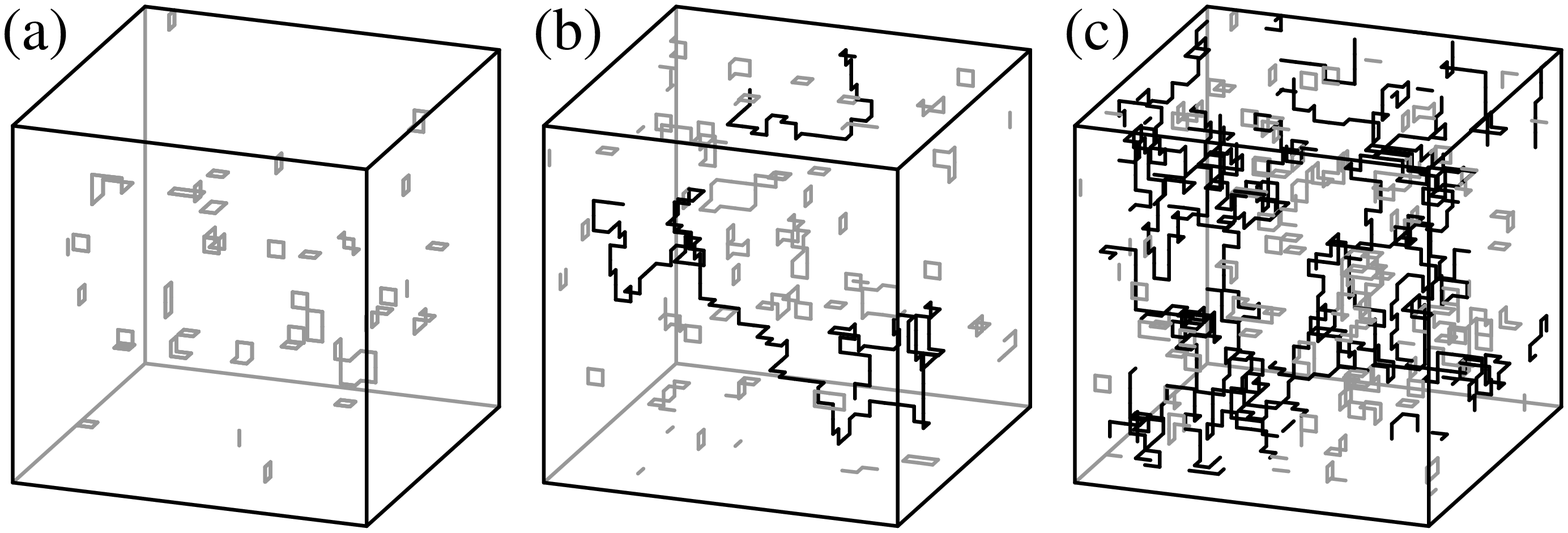}}
\caption{
Samples of loop configurations for a $3d$ hypercubic lattice 
with side length $L\!=\!24$ and periodic boundary conditions.
Percolating (nonpercolating) loops are colored black (gray).
The snapshots relate to different values of the
disorder parameter $\rho$, i.e.\ (a) $\rho\!=\!0.10$, 
(b) $\rho\!=\!0.13$, (c) $\rho\!=\!0.17$, so as to illustrate
the NWP of loops.
In the limit of large system sizes and above the critical point 
$\rho_c\!=\!0.1273(3)$, the lattice features system-spanning
loops of total negative weight.
\label{fig:samples3D}}
\end{figure}

Recently, we have introduced \cite{melchert2008}
\emph{negative-weight percolation} (NWP),
a problem with subtle differences as compared to other string-like percolation
problems.
In NWP, one considers a regular lattice graph with periodic boundary conditions
(BCs), where adjacent sites are joined by undirected edges. Weights are
assigned to the edges, representing quenched random variables drawn from 
a distribution that allows for edge weights of either sign. The details
of the weight distribution are further controlled by a tunable disorder 
parameter, see Sec.\ \ref{sect:model}. 
For a given realization of the disorder, one then computes a 
configuration of loops, i.e.\ closed paths on the lattice graph, 
such that the sum of the edge weights that build up the loops attains
an \emph{exact} minimum  
and is negative. Note that the application 
of exact algorithms in contrast
to standard sampling approaches like Monte Carlo simulations avoids
problems like equilibration. Also, since the algorithms run
in polynomial time, large instances can be solved.
As an additional optimization constraint we impose the condition that 
the loops are not allowed to intersect; consequently there is no definition
of clusters in the NWP model. 
Due to the fact that a loop is not allowed to intersect with itself or 
with other loops in its neighborhood, it exhibits an ``excluded volume''
quite similar to usual self-avoiding walks (SAWs) \cite{stauffer1994}.
The problem of finding these loops can be cast into a minimum-weight
path (MWP) problem, outlined below in more detail.
A pivotal observation is that, as a function of the disorder parameter,  
the NWP model features a disorder-driven, geometric phase transition.
In this regard, depending on the disorder parameter, one can identify two distinct
phases: 
(i)  a phase where the loops are ``small'', meaning that the 
linear extensions of  the loops are small in comparison to the system size, 
see Fig.\ \ref{fig:samples3D}(a). 
(ii) a phase where ``large'' loops exist that span the entire lattice, see Fig.\ \ref{fig:samples3D}(c).
Regarding these two phases and in the limit of large system sizes, there is a particular 
value of the disorder parameter at which system-spanning (i.e.\ percolating) loops appear 
for the first time, see Fig.\ \ref{fig:samples3D}(b).
     
Previously, we have investigated the NWP phenomenon for $2d$ lattice graphs \cite{melchert2008}
using finite-size scaling (FSS) analyses, where we characterized the underlying transition by 
means of a whole set of critical exponents. 
Considering different disorder distributions and lattice geometries, the exponents where found 
to be universal in $2d$ and clearly distinct from those describing other percolation phenomena.
In a subsequent study we investigated the effect of dilution on the critical 
properties of the $2d$ NWP phenomenon \cite{apolo2009}. Therein we performed
FSS analyses to probe critical points along the critical line in the disorder-dilution plane 
that separates domains that allow/disallow system-spanning loops. 
One conclusion of that study was that bond dilution changes the universality class 
of the NWP problem. Further we found that, for bond-diluted lattices prepared
at the percolation threshold of $2d$ random percolation and at full disorder, the 
geometric properties of the system-spanning loops compare well to those of ordinary 
self-avoiding walks.

Here, we study the negative weight percolation problem on hypercubic 
lattice graphs in dimensions $d\!=\!2$ through $7$.  
The aim of the present
study is to determine the upper critical dimension of the NWP problem from 
computer simulations for systems with finite size. 
In this regard, we compute the ground state (GS) loop configurations for the NWP 
model for a fairly general disorder distribution (described below in Sec.\ \ref{sect:model}) 
and characterize the resulting loops using observables from percolation theory. 
We perform finite-size scaling analyses to extrapolate the results to the 
thermodynamic limit.
As a fundamental observable that provides information on whether the upper critical 
dimension $d_u$ is reached, we monitor the fractal dimension $d_f$ of the loops. 
The fractal dimension can be defined from the scaling of the average length $\langle \ell \rangle$
of the percolating loops as a function of system size $L$ according to $\langle \ell \rangle\!\sim\!L^{d_f}$.
In $2d$ we previously obtained the estimate $d_f\!=\!1.266(2)$ \cite{melchert2008}. 
This tells that in $2d$ the loops are, in a statistical sense, somewhat less convoluted than
SAWs ($d_f^{\rm SAW}\!=\!1.333$). 
For $d\!\geq\!d_u$ we expect to observe $d_f\!=\!2$, as for usual self-avoiding lattice
curves. This means, the ``excluded volume'' effect mentioned earlier becomes irrelevant and
the loops exhibit the same scaling as ordinary random walks. 

The remainder of the presented article is organized as follows.
In section \ref{sect:model}, we introduce the model in 
more detail and we outline the algorithm used to compute the loop 
configurations. In section \ref{sect:results}, we list the results of 
our numerical simulations and in section \ref{sect:conclusions} we 
conclude with a summary.


\section{Model and Algorithm\label{sect:model}}
In the remainder of this article we consider regular hypercubic lattice graphs
$G\!=\!(V,E)$ with side length $L$ and fully
periodic boundary conditions (BCs) in dimensions $d\!=\!2\ldots7$.
The considered graphs have $N\!=\!|V| \!=\!L^d$ sites
$i\!\in\!V$ and a number of $|E| \!=\!d N$ undirected edges  
$\{i,j\}\!\in\!E$ that join adjacent sites $i,j\!\in\!V$.  
We further assign a weight $\omega_{ij}$ to each edge contained in $E$, 
representing quenched random variables that introduce disorder to the lattice.
In the present work we consider lattices which exhibit a fraction
$(1-\rho)$ of edges with weight 1 and a fraction $\rho$ of edges
following a Gaussian disorder, i.e.,
\begin{equation}
P(\omega)=\rho \exp{(-\omega^2/2)}/\sqrt{2\pi} + (1-\rho) \delta(\omega-1)\,. 
\label{eq:disordeeq}
\end{equation} 
This allows explicitly
for loops $\mathcal{L}$ with a negative total weight
$\omega_{\mathcal{L}}\!=\!\sum_{\{i,j\}\in\mathcal{L}}\omega_{ij}$.
To support intuition: For any nonzero value of the disorder parameter $\rho$, a sufficiently
large lattice will exhibit at least ``small'' loops that have negative weight,
see Fig.\ \ref{fig:samples3D}(a). If the disorder parameter is large enough,
system-spanning loops with negative weight will exist, 
see Figs.\ \ref{fig:samples3D}(b),(c). 

The NWP problem statement then reads as follows:
Given $G$ together with a realization of the disorder,
determine a set $\mathcal{C}$ of loops such that the
configurational energy, defined as the sum of all the loop-weights
$\mathcal{E}\!=\!\sum_{\mathcal{L} \in \mathcal{C}} \omega_{\mathcal{L}}$, is
minimized. Therein, the  weight of an individual loop
is smaller than zero.
As further optimization constraint, the loops are not
allowed to intersect.  Note that $\mathcal{C}$ may also be empty 
(clearly this is the case for $\rho\!=\!0$).
The set of optimum
loops  is obtained using an appropriate transformation of
the original graph as detailed in \cite{ahuja1993}.  For the
transformed graphs, minimum-weight perfect matchings (MWPM) 
\cite{cook1999,opt-phys2001,melchertThesis2009} are calculated,
yielding  the loops for each particular instance  of the
disorder. This procedure allows for an efficient implementation
\cite{practicalGuide2009} of the simulation algorithms.
Here, we 
give a brief description of the algorithmic procedure that yields a 
minimum-weight set of loops for a given realization of the disorder. 
Fig.\ \ref{fig2abcd} illustrates the 3 basic steps, 
which are detailed next:

\begin{figure}[t!]
\centerline{
\includegraphics[width=1.0\linewidth]{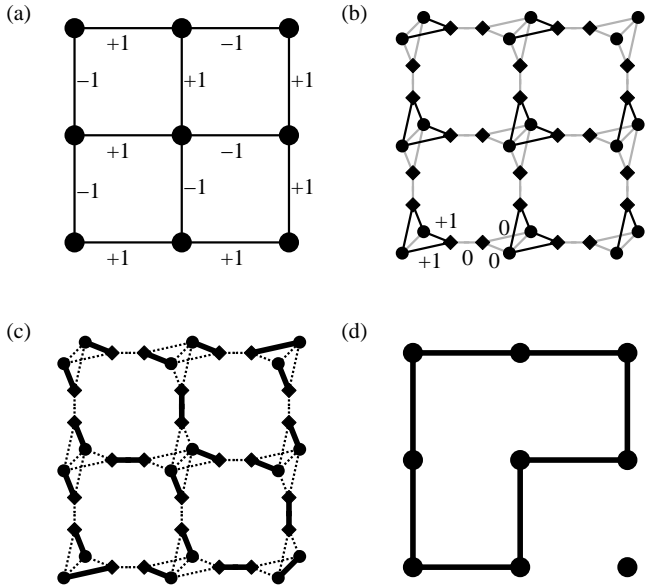}}
\caption{Illustration of the algorithmic procedure:
(a) original lattice $G$ with edge weights, 
(b) auxiliary graph $G_{\rm A}$ with proper weight assignment. Black 
edges carry the same weight as the respective edge in the original
graph and grey edges carry zero weight,
(c) minimum-weight perfect matching (MWPM) $M$: bold edges are matched 
and dashed edges are unmatched, and
(d) loop configuration (bold edges) that corresponds to the MWPM 
depicted in (c).
\label{fig2abcd}}
\end{figure}  

(1) each edge, joining adjacent sites on the original graph $G$,  is
replaced by a path of 3 edges.  Therefore, 2  ``additional'' sites
have to be introduced for each edge in $E$.  Therein, one of
the two edges connecting an additional site to an original site gets
the same weight as the corresponding edge in $G$. The remaining  two
edges get zero weight.  The original sites $i\in V$ are then
``duplicated'',  i.e. $i \rightarrow i_{1}, i_{2}$, along with all
their incident edges and the corresponding weights. 
 For each of these pairs of duplicated sites,
one additional  edge $\{i_1,i_2\}$ with zero weight is added that
connects the two sites $i_1$ and $i_2$.  The resulting auxiliary graph
$G_{{\rm A}}=(V_{{\rm A}},E_{{\rm A}})$  is shown
in Fig.\ \ref{fig2abcd}(b), where additional sites appear as squares and
duplicated  sites as circles. Fig.\ \ref{fig2abcd}(b) also illustrates
the weight  assignment on the transformed graph $G_{{\rm A}}$.  Note
that while the original graph (Fig.\ \ref{fig2abcd}(a)) is symmetric, the
transformed graph (Fig.\ \ref{fig2abcd}(b)) is not. This is due to the
details of the mapping procedure and the particular weight assignment
we have chosen.  A more extensive description of the mapping can be
found in \cite{melchert2007}.

(2) a MWPM on the auxiliary graph is determined via exact
combinatorial optimization algorithms \cite{comment_cookrohe}.
  A MWPM
is a minimum-weighted subset $M$ of $E_{\rm A}$, such that
each site  contained in $V_{\rm A}$ is met by precisely one
edge in $M$.  This is illustrated in Fig.\ \ref{fig2abcd}(c), where the
solid edges  represent $M$ for the given weight assignment. The dashed
edges are  not matched.  Due to construction, the auxiliary graph
consists of an even number of sites  and since there are no isolated
sites, it is guaranteed that a perfect matching exists.  \\
Note that obtaining the MWPM can be done in polynomial time
as a function of the number of sites, hence large system sizes with hundreds of
thousands of sites are feasible.

(3) finally it is possible to find a relation between the matched
edges $M$  on $G_{\rm A}$ and a configuration of negative-weighted
loops  $\mathcal{C}$ on $G$ by  tracing back the steps of the
transformation (1). Regarding this, note that each edge  contained
in $M$ that connects an additional site (square) to a duplicated  site
(circle) corresponds to an edge on $G$ that is part of a loop, see
Fig.\ \ref{fig2abcd}(d). 
Note that, by construction of the auxiliary graph,
 for each site $i_1$ or $i_2$ matched in this
way, the corresponding twin site $i_2$/$i_1$ must be matched
to an additional site as well. This guarantees that wherever a path enters
a site of the original graph, the paths also leaves the site, corresponding
to the defining condition of loops.
 All the edges in
$M$ that connect like  sites (i.e.\ duplicated-duplicated, or
additional-additional)  carry zero weight and do not contribute to a
loop on $G$.  Once the set $\mathcal{C}$ of loops is found, a
depth-first search \cite{ahuja1993,opt-phys2001} can  be used to
identify the loop set $\mathcal{C}$ and to determine the geometric 
properties of the individual loops. For the weight assignment illustrated 
in Fig.\ \ref{fig2abcd}(a), there is only one negative weighted loop with
$\omega_{\mathcal L}\!=\!-2$ and length $\ell=8$.

Note that the result of the calculation is a collection $\mathcal{C}$
of loops such that the total loop weight, and consequently the
configurational energy $\mathcal{E}$, is minimized. 
Hence, one obtains a global collective optimum of the system.  
Obviously, all loops that contribute to $\mathcal{C}$ possess a negative weight.  
Also note that the choice of the weight assignment in step (1) is not unique, 
i.e.\ there are different ways to choose a weight assignment
that all result in equivalent sets of matched edges on the transformed
lattice, corresponding to the minimum-weight collection of loops on
the original lattice. Some of these weight assignments result in a more
symmetric  transformed graph, see e.g. \cite{ahuja1993}. However, this
is only a technical issue that does not affect the resulting loop
configuration. Albeit the  transformed graph is not symmetric, the
resulting graph (Fig.\ \ref{fig2abcd}(d)) is again symmetric.
The small $2d$ lattice graph with free BCs shown in Fig.\ \ref{fig2abcd} was 
chosen intentionally for illustration purposes.
The algorithmic procedure extends 
to higher dimensions and fully periodic BCs in a straight-forward manner.

In the following section we will use the procedure outlined above to 
investigate the NWP phenomenon on hypercubic lattices.

\section{Results \label{sect:results}}
In the current section we will present the results of our simulations, 
carried out in order to characterize the critical behavior of the NWP phenomenon 
in dimensions $d\!=\!2\ldots7$.
To accomplish this, we use observables similar to those used 
in percolation theory and perform FSS analyses.  
The fundamental observables related to an individual loop $\mathcal{L}$
are its 
weight $\omega_{\mathcal{L}}$ and length $\ell=\sum_{\{i,j\}\in\mathcal{L}}1$.
Further, we determine the linear extensions $R_i$, $i\!=\!1\ldots d$, of a given loop 
by projecting it onto the independent lattice axes. The largest of those values, 
i.e.\ $R\!=\!\max_{i\!=\!1\ldots d}(R_i)$, is referred to as the spanning length 
of the loop.
To characterize the full perimeter of an individual loop 
on a coarse grained scale, we can further define the ``size'' $R_s\!=\!\sum_{i=1}^{d} R_i$, 
i.e.\ the length of the loop if all small scale irregularities where flattened \cite{vachaspati1984}.
The remainder of the present section is organized as follows.
In subsections \ref{subsect:Results_CritPoints} and \ref{subsect:Results_loopRatio}, 
we will first locate the critical points and exponents that characterize the NWP 
phenomenon on hypercubic lattice graphs. Therefore we perform FSS
analyses that involve data for different values of the disorder parameter. 
For these scaling analyses we considered hypercubic lattices with side lengths
up to $L_{\rm max}$, and a respective number of disorder configurations $n_{\rm max}$, as listed in Tab.\ \ref{tab:tab2}.
In subsection \ref{subsect:Results_crit} we will then state our results on the critical 
behavior of energetic and geometric loop-observables. Therefore, right at the critical 
points for the various dimensions, we perform simulations for lattices up to
$L_{\rm max}^{\rho_c}$ and $n_{\rm max}^{\rho_c}$, as listed in Tab.\ \ref{tab:tab2}.
%
\begin{table}[b!]
\caption{\label{tab:tab2}
System sizes and number of disorder configurations considered. From left to right:
dimension $d$, largest system size $L_{\rm max}$ and respective
number of samples $n_{\rm max}$ considered for the scaling analysis that involves
various values of the disorder parameter $\rho$, largest 
system size $L_{\rm max}^{\rho_c}$ and number of samples $n_{\rm max}^{\rho_c}$ considered for the 
analysis at $\rho_c$, and, number $N_{\rm loops}$ of loops collected at $L_{\rm max}^{\rho_c}$.
} 
\begin{ruledtabular}
\begin{tabular}[c]{l@{\quad}rr@{\quad}rr@{\quad}r}
$d$  & $L_{\rm max}$ & $n_{\rm max}$ & $L_{\rm max}^{\rho_c}$ & $n_{\rm max}^{\rho_c}$ & $N_{\rm loops}$\\ 
\hline
$2$ & $128$ & $40\,000$ & $512$ ($384$) & $3\,200$ ($21\,200$) & $25\,144\,685$\\ 
$3$ & $48$  & $9\,600$  & $56$  & $19\,200$ & $14\,292\,489$\\ 
$4$ & $24$  & $4\,800$  & $24$  & $9\,600$ & $4\,172\,813$\\ 
$5$ & $12$  & $6\,400$  & $12$  & $12\,200$ & $1\,762\,955$ \\ 
$6$ & $8$   & $6\,400$  & $8$   & $6\,400$ & $520\,368$ \\ 
$7$ & $5$   & $4\,800$  & $5$   & $12\,800$ & $204\,459$ \\
\end{tabular}
\end{ruledtabular}
\end{table}

\subsection{Scaling analyses to obtain critical points and exponents in $d\!=\!2\ldots7$} \label{subsect:Results_CritPoints}
\begin{figure}[t!]
\centerline{
\includegraphics[width=1.0\linewidth]{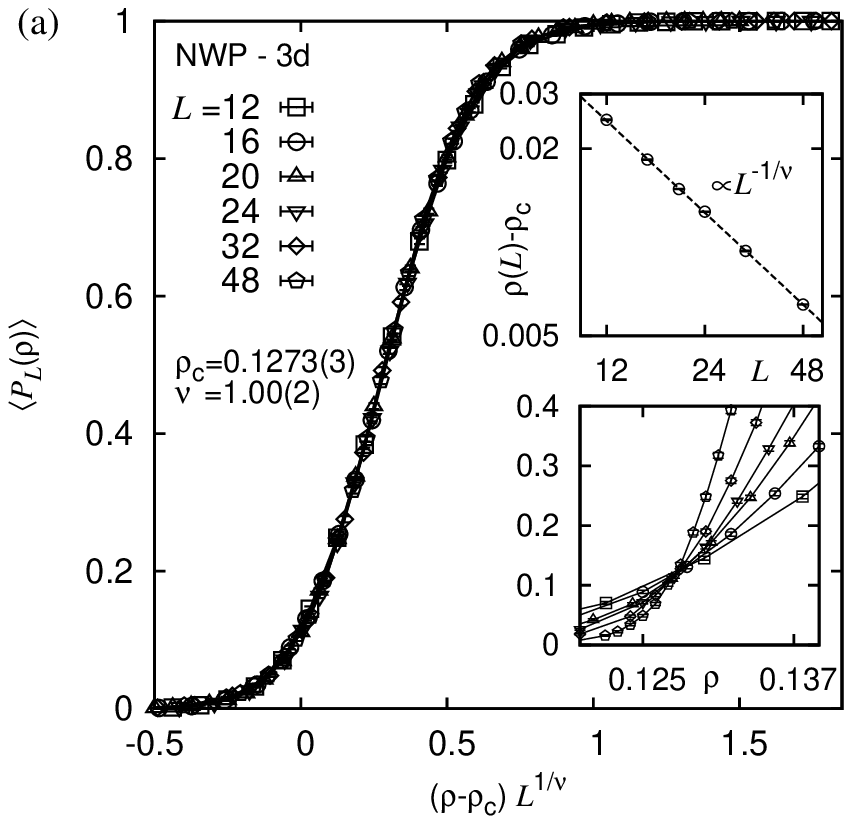}}
\centerline{
\includegraphics[width=1.0\linewidth]{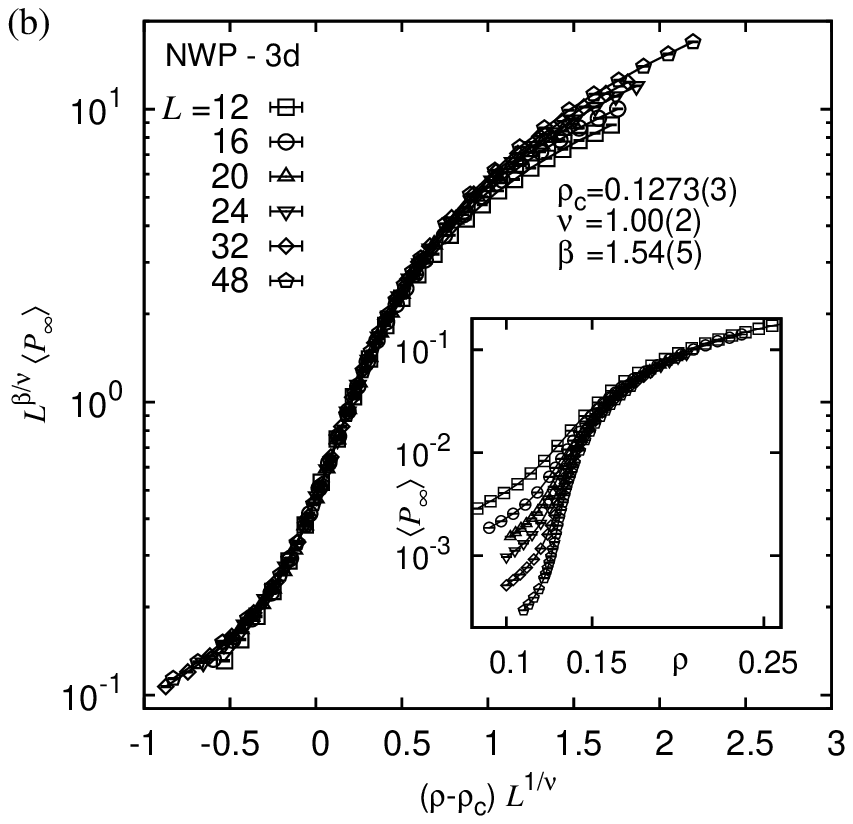}}
\caption{ \label{fig:fig3}
Results of the FSS analyses for the NWP problem 
on $3d$ hypercubic lattice graphs.
(a) scaling plot of the percolation probability $P_L(\rho)$. 
    The main plot shows the data collapse after rescaling
    the raw data according to Eq.\ \ref{eq:percProb}. 
    The inset at the bottom illustrates the unscaled data 
    close to the critical point $\rho_c$.
    The inset on top shows the scaling of the effective
    critical points $\rho(L)$, obtained from the finite 
    size fluctuations ${\rm var}(P_L(\rho))$.
(b) scaling plot of the order parameter $P_\infty$.
    The main plot shows the data collapse after rescaling
    the raw data according to Eq.\ \ref{eq:orderParam},
    and the inset shows the unscaled data.
}
\end{figure}

In the present subsection, we illustrate the analysis for the 
simulated data on $3d$ hypercubic lattices in detail. 
Although we performed similar
analyses for the remaining dimensions, we do not show figures
for $d\!=\!2,4-7$ but include the final results in Tab.\ \ref{tab:tab1}.
%
\begin{table}[b!]
\caption{\label{tab:tab1}
Critical properties that characterize the NWP phenomenon 
in $d\!=\!2\ldots7$. 
From left to right: 
Lattice dimension $d$, critical point $\rho_c$, critical exponent $\nu$
that describes the divergence of a typical length scale, 
order parameter exponent $\beta$, 
fluctuation exponent $\gamma$, fractal dimension $d_f$ at
criticality and Fisher exponent $\tau$. Note that the figures for $d\!=\!2$ 
are taken from Ref.\ \cite{melchert2008}.
} 
\begin{ruledtabular}
\begin{tabular}[c]{l@{\quad}llllll}
$d$  & $\rho_c$ & $\nu$ & $\beta$ & $\gamma$ & $d_f$ & $\tau$ \\
\hline
2 & 0.340(1)  & 1.49(7) & 1.07(6) &  0.77(7) & 1.266(2) & 2.59(3)\\
3 & 0.1273(3) & 1.00(2) & 1.54(5) & -0.09(3) & 1.459(3)& 3.07(1)\\
4 & 0.0640(2) & 0.80(3) & 1.91(11)& -0.66(5) & 1.60(1) & 3.55(2)\\
5 & 0.0385(2) & 0.66(2) & 2.10(12)& -1.06(7) & 1.75(3) & 3.86(3)\\
6 & 0.0265(2) & 0.50(1) & 1.92(6) & -0.99(3) & 2.00(1) & 4.00(2)\\ 
7 & 0.0198(1) & 0.41(1) & --      & --       & 2.08(8) & 4.50(1) 
\end{tabular}
\end{ruledtabular}
\end{table}
As pointed out earlier, a loop is called percolating if its spanning length $R$ is equal 
to the system size $L$. This is a binary decision for each realization of the disorder and it is further used
to obtain the percolation probability $P_L(\rho)$ for a lattice graph of
a certain size $L$ at a given value of the disorder parameter $\rho$. 
According to scaling theory, one expects $P_L(\rho)$ to satisfy the scaling expression
\begin{eqnarray}
\langle P_L(\rho) \rangle \sim f_1[(\rho-\rho_c)L^{1/\nu}], \label{eq:percProb}
\end{eqnarray}
wherein $\langle \ldots \rangle$ denotes the disorder average and
$\rho_c$ is the critical value of the disorder parameter above which system
spanning loops first appear as $L\!\to\!\infty$. 
Further, $\nu$ is a critical exponent that
describes the divergence of a typical length scale in the 
NWP problem as the critical point is approached. Finally, $f_1[\cdot]$
denotes an (unknown) universal scaling function.
Eq.\ \ref{eq:percProb} implies that if one plots $P_L(\rho)$ as a function of the 
scaled variable $x\equiv (\rho-\rho_c)L^{1/\nu}$ and if one adjusts $\rho_c$ and $\nu$
to their proper values, one should find a collapse of the data curves belonging to different 
values of $L$ onto a master curve.
Note that above, $x$ constitutes a lowest order polynomial approximation 
to $f_1[x]$ regarding the disorder parameter $\rho$ around the critical point $\rho_c$.
The resulting scaling plot for the data of $3d$ hypercubic lattices is shown in 
Fig.\ \ref{fig:fig3}(a). Therein, considering Eq.\ \ref{eq:percProb}, a best data
collapse of the curves for $L\!\geq\!24$ yields the parameters
$\rho_c\!=\!0.1273(3)$ and $\nu\!=\!1.00(2)$ ($S\!=\!1.02$),
where the scaling analysis was restricted to the finite interval $dx\!=\![-0.2\!:\!0.4]$ enclosing
the critical point on the rescaled $x$-axis.
The value of $S$ measures the mean square deviation of the data points from the master curve
in units of the standard error and thus provides information on how well the simulated data
fits the scaling expression, see Refs.\ \cite{houdayer2004,autoScale2009}. Here, the 
data collapse is considered to be good if the numerical value of $S\!\leq\!2$.
Further, the quality $S$ of the data collapse and the resulting estimates for the 
critical parameters did not depend much on the size of the chosen interval $dx$.
As an alternative, the maxima of the associated fluctuations, i.e.\ 
${\rm var}(P_L(\rho))\!=\!\langle P_L(\rho)^2\rangle \!-\!\langle P_L(\rho)\rangle^2$,
can be used to define system size dependent, ``effective'' critical points $\rho(L)$ \cite{stauffer1994}. 
These maxima are located at precisely those values of $\rho$ where 
$P_L(\rho)\!=\!1/2$, and just as $P_L(\rho)$ approaches a step function in the thermodynamic 
limit, $\rho(L)$ approaches $\rho_c$ as $L\!\to\!\infty$. 
In this regard, we expect the sequence of effective critical points to attain an asymptotic
value as $\rho(L)\!=\!\rho_c+aL^{-1/\nu}$. First, we obtained the estimates of $\rho(L)$ by fitting a 
Gaussian function to the peaks of ${\rm var}(P_L(\rho))$. Applying the above scaling form
to the data points thus obtained (see upper inset of Fig.\ \ref{fig:fig3}(a)), then yields 
$\rho_c\!=\!0.1270(4)$ and $\nu\!=\!1.02(4)$ in agreement with the estimates reported earlier.
Further, for each realization of the disorder we can compute the size of the smallest
box that fits the largest loop on the lattice, i.e.\ $V_{\rm B}\!=\!\Pi_{i=1}^{d}R_i$.
For the normalized box-size we observe the scaling behavior 
$\langle V_{\rm B}/L^d\rangle\!\sim\!f_2[(\rho-\rho_c)L^{1/\nu}]$ (not shown), with scaling parameters 
$\rho_c\!=\!0.1273(2)$ and $\nu\!=\!0.99(4)$ ($S\!=\!1.00$).
Since the analyses related to these three different observables conclude
with scaling parameters that agree within the error bars, we are confident
that the respective values of $\rho_c$ and $\nu$, listed in Tab.\ \ref{tab:tab1}, 
properly describe the critical behavior of the NWP phenomenon on $3d$ hypercubic 
lattice graphs.

A second critical exponent is related to the scaling behavior of the order parameter
$P_\infty\!=\!\ell/L^d$, which measures the probability that a site on the 
lattice graph belongs to the largest loop. Therein, $\ell$ refers to the length of the largest 
loop for each realization of the disorder. According to scaling
theory one can expect $P_\infty$ to scale as 
\begin{eqnarray}
\langle P_\infty \rangle \sim L^{-\beta/\nu}f_3[(\rho-\rho_c)L^{1/\nu}], \label{eq:orderParam}
\end{eqnarray}
wherein $\beta$ signifies the order parameter exponent.
Again, for the $3d$ data, a FSS analysis utilizing a collapse of the
data curves for $L\!\geq\!24$ yields the estimate $\beta\!=\!1.54(5)$ ($S\!=\!1.24$).
A scaling plot of the order parameter is presented in Fig.\ \ref{fig:fig3}(b). 
Therein, the data collapse is best close to the critical point. So as to reduce the effect of 
the corrections to scaling off criticality, the scaling analysis was restricted to the 
finite interval $dx\!=\![-0.15\!:\!0.225]$ on the rescaled $x$-axis.

The corresponding estimates of $\rho_c$, $\nu$ and $\beta$ for hypercubic lattice 
graphs in $d\!=\!2,4-7$, resulting from similar FSS analyses, are listed in Tab.\ \ref{tab:tab1}.

\subsection{Scaling analysis of the loop-length ratio}\label{subsect:Results_loopRatio}

During the simulations we recorded the energetic and geometric properties of the largest 
and $2$nd largest loops, with respective lengths $\ell_1$ and $\ell_2$, for each realization 
of the disorder. 
The average loop-length ratio $\langle \ell_1/\ell_2 \rangle$ for these loops was found to 
satisfy the scaling expression 
\begin{eqnarray}
\langle \ell_1/\ell_2\rangle \sim f_4[(\rho-\rho_c)L^{1/\nu}]. \label{eq:lenRatio}
\end{eqnarray}
In order to assess the corresponding scaling behavior, we discarded samples 
that featured less than two loops (i.e.\ samples with $\ell_2\!=\!0$).
A similar scaling for the cluster-size ratio was previously confirmed for usual random 
percolation \cite{dasilva2002}. It stems from the fact that the largest and $2$nd largest clusters
exhibit the same fractal dimension at the critical point.
For usual percolation this issue was addressed earlier \cite{jan1998}.  
Further, we observed a similar scaling behavior in the context of an analysis of
ferromagnetic spin domains at the $T\!=\!0$ spin glass to ferromagnet transition 
for the $2d$ random bond Ising model \cite{melchert2009}.
\begin{figure}[t!]
\centerline{
\includegraphics[width=1.0\linewidth]{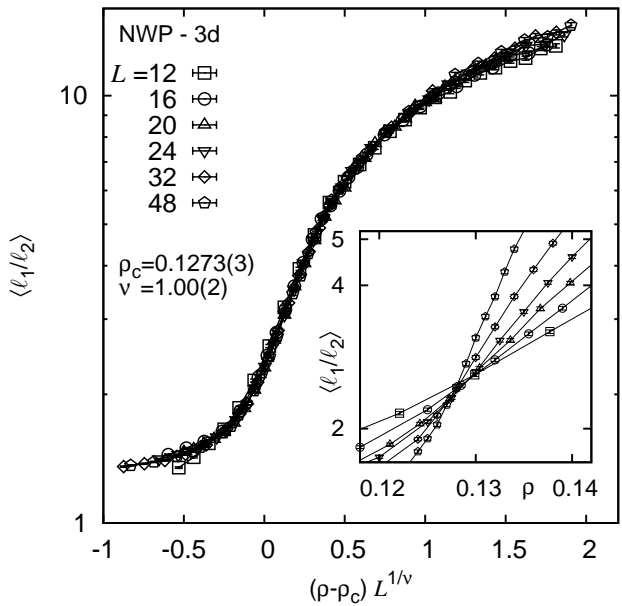}}
\caption{ \label{fig:fig4}
Results of the FSS analysis of the loop length ratio 
$\langle \ell_1/\ell_2\rangle$ on $3d$ hypercubic lattice graphs.
The main plot shows the data collapse after rescaling 
the simulated data according to Eq.\ \ref{eq:lenRatio} and the 
inset illustrates the unscaled data close to the critical point.
}
\end{figure}  

Regarding the data for hypercubic lattices of different dimensions $d$ and considering
Eq.\ \ref{eq:lenRatio}, we here yield the estimates 
\begin{align*}
3d&:  &\rho_c\!=& 0.1274(3) &\nu\!=& 0.99(5) &S\!=& 0.87 && [-0.45\!:\!0.45] \\
4d&:  && 0.0641(4) && 0.80(9) && 0.69 && [-0.30\!:\!0.35] \\
5d&:  && 0.0382(4) && 0.68(9) && 0.57 && [-0.75\!:\!0.35]\\
6d&:  && 0.0262(1) && 0.50(3) && 0.73 && [-0.13\!:\!0.28]
\end{align*}
that agree with those obtained earlier in subsection \ref{subsect:Results_CritPoints},
listed in Tab.\ \ref{tab:tab1}, within error bars. 
Note that for the $7d$ systems,
our data did not allow for a decent analysis of the loop-length ratio.
Also, there are no results listed for the $2d$ case. This is so, since at 
the time we performed the simulations for the $2d$ square systems, we did not
write out the second largest loop length, explicitly.
A scaling plot that illustrates the behavior of the loop length ratio for the $3d$ systems
is presented as Fig.\ \ref{fig:fig4}. 
Further, note that the estimates of the scaling parameters (for the various values of $d$) 
did not depend much on the size of the considered scaling interval. 
E.g., for the $3d$ systems considering $dx\!=\![-0.4:1.25]$, we obtained $\rho_c\!=\!0.1273(4)$ 
and $\nu\!=\!1.00(6)$ with the somewhat larger quality $S\!=\!0.97$.

Note that the scaling according to Eq.\ \ref{eq:lenRatio} was established empirically. 
So as to check whether that scaling assumption fits the data well, we allowed for a further 
free parameter, considering a scaling of the form 
$\langle \ell_1/\ell_2\rangle\!\sim\!L^{\kappa}f_5[(\rho-\rho_c)L^{1/\nu}]$.
We found that the best data collapse for given intervals $dx$ where attained for 
values $\rho_c$ and $\nu$ in agreement with those above and $|\kappa|\!\approx\!10^{-3}$.

\subsection{Scaling at the critical point}\label{subsect:Results_crit}

As pointed out above, during the simulation we recorded the linear extensions 
$R_i$, $i\!=\!1\ldots d$, of the individual loops by projecting it onto the independent 
lattice axes. 
So as to study the scaling of the loop shape,
we collected, for each dimension $d$, a large number $N_{\rm loops}$ 
of loops (see Tab.\ \ref{tab:tab2}) at the critical point $\rho_c$ for 
the largest system size $L_{\rm max}^{\rho_c}$ considered for the respective setup.
For those loops we then monitored the volume to surface ratio $V_{\rm B}/S_{\rm B}$ of 
the smallest box that fits the individual loops as a function 
of the coarse-grained 
loop size $R_s=\sum_{i=1}^d R_i$, where $V_{\rm B}\!=\!\Pi_{i=1}^{d}R_i$ 
and $S_{\rm B}\!=\!2\times\sum_{i=1}^{d} V_{\rm B}/R_i$.
For hypercubic volumes with identical values $R_i$, $i\!=\!1\ldots d$, one would expect to find 
$V_{\rm B}/S_{\rm B}\!=\!(2 d d)^{-1}R_s$. Considering $d\!=\!2\dots 6$ and performing 
fits to the form $\langle V_{\rm B}/S_{\rm B}\rangle_{R_s}\!=\!cR_s^{\psi}$ we yield 
$|\psi-1|\!\approx\!10^{-2}$ and values of $c$ reasonably close to $(2 d d)^{-1}$ in order 
to conclude that the loops, in a statistical sense, are not oblate but possess a rather spherical shape.
E.g., in $3d$ we obtained $c/(2 d d)\!=\!0.95(1)$ and $\psi\!=\!1.00(3)$.
However, in $7d$ the data is not well represented by the scaling form above. 
In this regard, we found our data best fit by the precise scaling form
$\langle V_{\rm B}/S_{\rm B}\rangle_{\ell}\!=\! 0.003(1)\!(\ell\!+\!15(3))^{1.1(1)}$, where we 
considered the ``true'' loop length $\ell$ instead of $R_s$. 
Unfortunately, this contains no information that relates to the 
``loop-shape factor'' $(2 d d)^{-1}$ introduced above.

\begin{figure}[t!]
\centerline{
\includegraphics[width=1.0\linewidth]{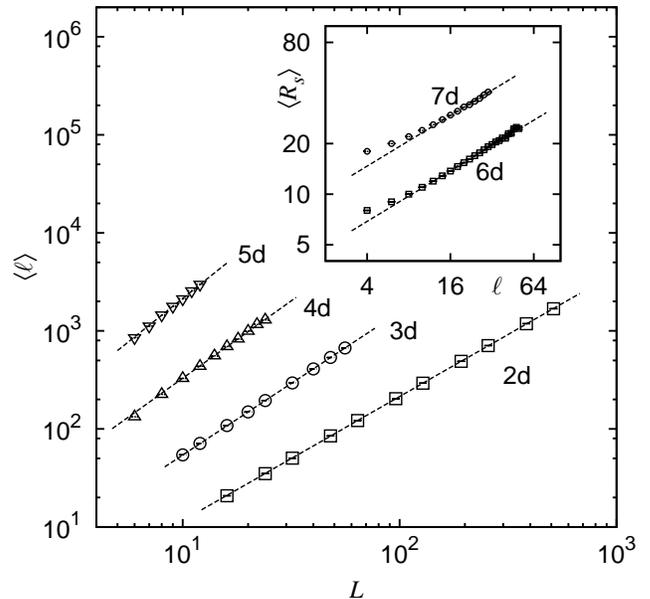}}
\caption{ \label{fig:fig5}
Results of the FSS analyses to estimate the fractal dimension
$d_f$ of the loops. The main plot shows the scaling of 
the average loop length $\langle \ell \rangle$ as function 
of the system size $L$ for $d\!=\!2\ldots5$. The dashed lines
indicate the asymptotic scaling according to 
$\langle \ell \rangle\!\sim\!L^{d_f}$, with values $d_f$ listed in 
Tab.\ \ref{tab:tab1}. 
Note that the data sets where shifted upwards by a factor $4$, $20$ and
$100$ for $d\!=\!3$, $4$, $5$, respectively.
The inset shows the scaling of the average loop size $\langle R_s \rangle$
as function of the ``true'' loop length $\ell$ for $d\!=6,7$. 
The dashed lines are $\sim\!\ell^{1/2}$, to which the asymptotic scaling
$\langle R_s \rangle\!\sim\!\ell^{1/d_f}$ can be compared. 
The $7d$ data was shifted upwards by a factor $2$.
}
\end{figure}  
Next, we aim to determine the fractal dimension $d_f$ of the loops, which can be defined from the
scaling behavior of the average loop length $\langle \ell \rangle$ as a function of the linear 
extend $L$ of the hypercubic lattice graphs at the critical point $\rho_c$ according to 
$\langle \ell \rangle \!\sim\! L^{d_f}$.
For dimensions $d\!=\!2\ldots5$ we thus analyzed the largest loop found for each realization 
of the disorder and employed the scaling relation above, see Fig.\ \ref{fig:fig5}. The resulting estimates
for $d_f$ are listed in Tab.\ \ref{tab:tab1}.
Further, we verified that the probability density function $D_L(\rho)$ of the largest loop length found
for each realization of the disorder yields a data collapse after a rescaling of the form 
$D_L(\ell)\!\sim\!L^{-d_f}f_6[\ell/L^{d_f}]$ (not shown).
Due to the few and small system sizes that can be reached in $d\!=\!6$, $7$ ($L_{\rm max}^{\rho_c}\!=\!8,5$, respectively), 
the data analysis turned out to be somewhat more intricate. 
For those two cases we considered only the largest lattice and analyzed the scaling behavior of all the small, i.e.\
nonpercolating, loops, where we considered the scaling form $\langle R_s \rangle\!\sim\!\ell ^{1/d_f}$. 
For the considered lattice sizes the values of $\ell$ where not too 
diverse and we collected $520368$ ($6d$) and $204459$ ($7d$) loops that comprise the estimates 
$d_f\!=\!2.00(1)$ ($6d$) and $2.08(8)$ ($7d$), see Fig.\ \ref{fig:fig5}. 
However, note that for the data analysis all those data points have to be discarded that are strongly affected by the
granularity of the lattice. For this reason, all the data points for $\ell\!\leq\!10$ have been withdrawn.
Unfortunately, at $\rho_c$, the number density $n_{\ell}$ of loops with a given length $\ell$ decays algebraically 
as $n_{\ell}\!\sim\!\ell^{-\tau}$, where $\tau\!\geq\!1+d/2$ (see below).
This means, considering $\ell\!>\!10$, the values of $d_f$ obtained from the scaling form above stem 
from only a fraction of the collected loops.
E.g., for $6d$ and $\ell\!>\!10$ we have only $5884$ loops that represent the respective averages $\langle R_s \rangle$.
Hence, the results for $d\!=\!6$ and $7$ have to be taken with a grain of salt.
However, the fact that $d\!=\!6$ is the smallest dimension for which the fractal dimension of the
loops attains the value of $d_f\!=\!2$ suggests an upper critical dimension $d_u\!=\!6$ for the NWP phenomenon.
In a previous study \cite{melchert2008} we found that for $2d$ systems, 
the weight $\omega_{\mathcal{L}}$ of a loop $\mathcal{L}$ is proportional to its length $\ell$.
Here, we verified the same behavior for the various dimensions considered.
More precise, we collected loops for the largest system size $L_{\rm max}^{\rho_c}$ at the critical 
point $\rho_c$ of a given dimension $d$. Regarding the loop weight we found a best fit to 
the data by using the scaling form $\langle \omega_{\mathcal{L}}\rangle\!\sim\!\ell (1+c_1/\ell^{c_2})$,
wherein $c_1$ was of order $10$ and $c_2\!\approx\!1$ for all dimensions considered. 

Another critical exponent can be obtained from the scaling of the finite size susceptibility associated 
to the order parameter, i.e.\ $\chi_L\!=\!L^d {\rm var}(P_\infty)\!\equiv\!L^{-d}{\rm var}(\ell)$. 
Basically, this observable measures the mean-square fluctuation of the loop length and it exhibits
the critical scaling $\chi_L\!\sim\!L^{\gamma/\nu}$ (not shown). The resulting estimates of the fluctuation exponent
$\gamma$ are listed in Tab.\ \ref{tab:tab1} and are found to agree with the scaling relation $\gamma+2\beta=d \nu$ within 
error bars. Note that in $7d$ the quality of the data was not sufficient to obtain an estimate for $\gamma$. 

\begin{figure}[t!]
\centerline{
\includegraphics[width=1.0\linewidth]{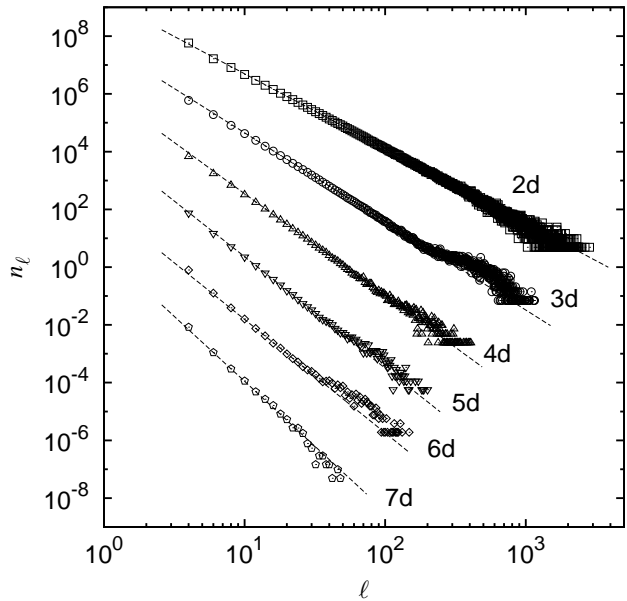}}
\caption{ \label{fig:fig6}
Results of the FSS analyses for the number density $n_\ell$ of 
all nonpercolating loops with length $\ell$ for $d\!=\!2\ldots7$.
Right at $\rho_c$, the number density exhibits an algebraic scaling
$n_\ell\!\sim\!\ell^{-\tau}$, with $\tau$ listed in Tab.\ \ref{tab:tab1}. 
Note that the the data sets for dimensions $d$ where scaled by 
a factor $10^{2(6-d)}$.
}
\end{figure}

Finally, we investigate the number density $n_\ell$ of all nonpercolating loops with length $\ell$.
Right at the critical point, it is expected to exhibit an algebraic
scaling $n_\ell\!\sim\!\ell^{-\tau}$, governed by the Fisher exponent $\tau$. 
For the largest lattice graphs simulated for the various values of $d$, we obtain
the estimates listed in Tab.\ \ref{tab:tab1}, see also Fig.\ \ref{fig:fig6}.
For the corresponding data analyses, very small loops have to be neglected since
they are affected by the granularity of the lattice and very large loops have 
to be withdrawn since they are affected by the lattice boundaries.
From scaling, the Fisher exponent can be related to the fractal dimension 
via the scaling relation $\tau-1\!=\!d/d_f$. Note that the values of $\tau$ 
and $d_f$ listed in Tab.\ \ref{tab:tab1} where obtained independently and
are found to agree with the latter scaling relation within error bars, in
support of the estimate $d_u\!=\!6$ suggested above. 

\section{Conclusions \label{sect:conclusions}}

In the present study, we performed numerical simulations on hypercubic lattice
graphs with ``Gaussian-like'' disorder in dimensions $d\!=\!2$ through $7$.
The aim of the study was to identify the upper critical dimension of the NWP phenomenon.
Therefore, we used a mapping of the NWP model to a combinatorial optimization problem
that allows to obtain configurations of minimum weight loops by means of exact algorithms.
We characterized the loops using observables from percolation theory and performed
FSS analyses to estimate critical points and exponents that describe the disorder
driven, geometric phase transition related to the NWP problem in the different dimensions.

Albeit the data analysis is notoriously difficult for large values of $d$, 
we find our results consistent with an upper critical dimension $d_u\!=\!6$ for the 
NWP model.
This conclusion was based on the estimates of the fractal dimension of the loops, which,
in $6d$ attains the value $d_f\!=\!2$ for the first time (bear in mind that $d_f\!=\!2$ indicates the scaling
of a completely uncorrelated lattice curve).
Further, in $6d$, the critical exponent $\nu\!=\!0.5$ that describes the divergence of a typical length
scale in the NWP problem matches the value of $\nu$ for usual random percolation at the
upper critical dimension \cite{stauffer1994}.
According to our results, the FSS exponent $\nu$ still changes for $d\!>\!d_u$, which, at a first
glance appears to be a little odd. However, this seems to be in agreement with the FSS for random 
percolation above six dimensions, where it was found that the corresponding 
exponent takes the value $3/d$ \cite{aharony1995}. Moreover, the value $\nu\!=\!0.41(1)$ 
for the $7d$ systems found here is close to the percolation estimate $3/7\!\approx\!0.429$.

At this point, we would like to note that its tempting to perform simulations for the NWP problem on 
random graphs, where one has direct access to the mean field exponents that describe the
transition. Since the upper critical dimension can be defined as the smallest dimension 
for which the critical exponents take their mean field values, such simulations could be used
to provide further support for the result $d_u\!=\!6$ obtained here.

Note that rather similar results where found in the context of the 
optimal-path 
problem \cite{schwartz1998}, wherein one aims to minimize the largest weight along a single path,
in contrast to minimizing the sum of weights of multiple loops, as above. 
Further, the optimal path problem can be mapped to the 
minimum-spanning tree problem 
\cite{dobrin2001} and to invasion percolation with trapping \cite{barabasi1996}.
Regarding the optimal path problem in strong disorder \cite{buldyrev2004}, 
quite similar fractal scaling exponents can be observed: $d_{\rm opt}\!=\!1.222(3)$ ($2d$, Ref.\ \cite{middleton2000}), 
$1.44(1)$ ($3d$, Ref.\ \cite{buldyrev2004}) and $1.59(2)$ ($4d$, Ref.\ \cite{cieplak1996} wherein also the approximate 
scaling relation $d_{\rm opt}\!=\!(d+4)/5$ was hypothesized).
The correspondence to invasion percolation with trapping further suggests an upper critical dimension 
$d^{\rm opt}_u\!=\!6$ \cite{buldyrev2004} for the optimal path problem.

Finally, we will elaborate on the results for the $3d$ systems.
In an earlier study \cite{melchert2008}, we performed simulations for $3d$ hypercubic
lattice graphs respecting a bimodal distribution ($\omega\!=\!\pm 1$) of the edge-weights. 
Therein, the most reliable results include the estimates $\nu\!=\!1.02(3)$, obtained from a
FSS analysis of the percolation probability, and $d_f\!=\!1.43(2)$, obtained 
from the scaling of the ``small'' loops at the respective critical point.
These values are reasonably close to those found here for the ``Gaussian-like''
disorder in order to conclude that the exponents in $3d$ are universal, i.e.\
they do not depend on minor details of the problem setup as, e.g., the disorder distribution.
Further, the exponents $\nu$, $\beta$ and $d_f$ for the $3d$ setup found here are close by those that
describe the disorder induced vortex loop percolation transition for the superconductor-to-normal 
transition in a $3d$ strongly screened vortex glass model \cite{pfeiffer2002}.


\begin{acknowledgments}
LA  acknowledges a scholarship of the German academic exchange service DAAD
within the ``Research Internships in Science and Engineering'' (RISE) program
and the City College Fellowship program for further support.
We further acknowledge financial support from the VolkswagenStiftung (Germany)
within the program ``Nachwuchsgruppen an Universit\"aten''. The
simulations were performed at the GOLEM I cluster for scientific computing at 
the University of Oldenburg (Germany).
\end{acknowledgments}


\bibliography{nwp_upperCritDim.bib}

\begin{thebibliography}{33}
\expandafter\ifx\csname natexlab\endcsname\relax\def\natexlab#1{#1}\fi
\expandafter\ifx\csname bibnamefont\endcsname\relax
  \def\bibnamefont#1{#1}\fi
\expandafter\ifx\csname bibfnamefont\endcsname\relax
  \def\bibfnamefont#1{#1}\fi
\expandafter\ifx\csname citenamefont\endcsname\relax
  \def\citenamefont#1{#1}\fi
\expandafter\ifx\csname url\endcsname\relax
  \def\url#1{\texttt{#1}}\fi
\expandafter\ifx\csname urlprefix\endcsname\relax\def\urlprefix{URL }\fi
\providecommand{\bibinfo}[2]{#2}
\providecommand{\eprint}[2][]{\url{#2}}

\bibitem[{\citenamefont{Kremer}(1981)}]{kremer1981}
\bibinfo{author}{\bibfnamefont{K.}~\bibnamefont{Kremer}}, \bibinfo{journal}{Z.
  Phys. B} \textbf{\bibinfo{volume}{45}}, \bibinfo{pages}{149}
  (\bibinfo{year}{1981}).

\bibitem[{\citenamefont{Kardar and Zhang}(1987)}]{kardar1987}
\bibinfo{author}{\bibfnamefont{M.}~\bibnamefont{Kardar}} \bibnamefont{and}
  \bibinfo{author}{\bibfnamefont{Y.~C.} \bibnamefont{Zhang}},
  \bibinfo{journal}{Phys. Rev. Lett.} \textbf{\bibinfo{volume}{58}},
  \bibinfo{pages}{2087} (\bibinfo{year}{1987}).

\bibitem[{\citenamefont{Derrida}(1990)}]{derrida1990}
\bibinfo{author}{\bibfnamefont{B.}~\bibnamefont{Derrida}},
  \bibinfo{journal}{Physica A} \textbf{\bibinfo{volume}{163}},
  \bibinfo{pages}{71} (\bibinfo{year}{1990}).

\bibitem[{\citenamefont{Grassberger}(1993)}]{grassberger1993}
\bibinfo{author}{\bibfnamefont{P.}~\bibnamefont{Grassberger}},
  \bibinfo{journal}{J. Phys. A} \textbf{\bibinfo{volume}{26}},
  \bibinfo{pages}{1023} (\bibinfo{year}{1993}).

\bibitem[{\citenamefont{Parshani et~al.}(2009)\citenamefont{Parshani,
  Braunstein, and Havlin}}]{parshani2009}
\bibinfo{author}{\bibfnamefont{R.}~\bibnamefont{Parshani}},
  \bibinfo{author}{\bibfnamefont{L.~A.} \bibnamefont{Braunstein}},
  \bibnamefont{and} \bibinfo{author}{\bibfnamefont{S.}~\bibnamefont{Havlin}},
  \bibinfo{journal}{Phys. Rev. E} \textbf{\bibinfo{volume}{79}},
  \bibinfo{pages}{050102} (\bibinfo{year}{2009}).

\bibitem[{\citenamefont{Pfeiffer and Rieger}(2002)}]{pfeiffer2002}
\bibinfo{author}{\bibfnamefont{F.~O.} \bibnamefont{Pfeiffer}} \bibnamefont{and}
  \bibinfo{author}{\bibfnamefont{H.}~\bibnamefont{Rieger}},
  \bibinfo{journal}{J. Phys.: Condens. Matter} \textbf{\bibinfo{volume}{14}},
  \bibinfo{pages}{2361} (\bibinfo{year}{2002}).

\bibitem[{\citenamefont{Pfeiffer and Rieger}(2003)}]{pfeiffer2003}
\bibinfo{author}{\bibfnamefont{F.~O.} \bibnamefont{Pfeiffer}} \bibnamefont{and}
  \bibinfo{author}{\bibfnamefont{H.}~\bibnamefont{Rieger}},
  \bibinfo{journal}{Phys. Rev. {\bf E}} \textbf{\bibinfo{volume}{67}},
  \bibinfo{pages}{056113} (\bibinfo{year}{2003}).

\bibitem[{\citenamefont{Cieplak et~al.}(1994)\citenamefont{Cieplak, Maritan,
  and Banavar}}]{cieplak1994}
\bibinfo{author}{\bibfnamefont{M.}~\bibnamefont{Cieplak}},
  \bibinfo{author}{\bibfnamefont{A.}~\bibnamefont{Maritan}}, \bibnamefont{and}
  \bibinfo{author}{\bibfnamefont{J.~R.} \bibnamefont{Banavar}},
  \bibinfo{journal}{Phys. Rev. Lett.} \textbf{\bibinfo{volume}{72}},
  \bibinfo{pages}{2320} (\bibinfo{year}{1994}).

\bibitem[{\citenamefont{Melchert and Hartmann}(2007)}]{melchert2007}
\bibinfo{author}{\bibfnamefont{O.}~\bibnamefont{Melchert}} \bibnamefont{and}
  \bibinfo{author}{\bibfnamefont{A.~K.} \bibnamefont{Hartmann}},
  \bibinfo{journal}{Phys. Rev. B} \textbf{\bibinfo{volume}{76}},
  \bibinfo{pages}{174411} (\bibinfo{year}{2007}).

\bibitem[{\citenamefont{Schwarz et~al.}(2009)\citenamefont{Schwarz,
  Karrenbauer, Schehr, and Rieger}}]{schwarz2009}
\bibinfo{author}{\bibfnamefont{K.}~\bibnamefont{Schwarz}},
  \bibinfo{author}{\bibfnamefont{A.}~\bibnamefont{Karrenbauer}},
  \bibinfo{author}{\bibfnamefont{G.}~\bibnamefont{Schehr}}, \bibnamefont{and}
  \bibinfo{author}{\bibfnamefont{H.}~\bibnamefont{Rieger}},
  \bibinfo{journal}{J. Stat. Mech.} \textbf{\bibinfo{volume}{2009}},
  \bibinfo{pages}{P08022} (\bibinfo{year}{2009}).

\bibitem[{\citenamefont{Stauffer}(1979)}]{stauffer1979}
\bibinfo{author}{\bibfnamefont{D.}~\bibnamefont{Stauffer}},
  \bibinfo{journal}{Phys. Rep.} \textbf{\bibinfo{volume}{54}},
  \bibinfo{pages}{1} (\bibinfo{year}{1979}).

\bibitem[{\citenamefont{Stauffer and Aharony}(1994)}]{stauffer1994}
\bibinfo{author}{\bibfnamefont{D.}~\bibnamefont{Stauffer}} \bibnamefont{and}
  \bibinfo{author}{\bibfnamefont{A.}~\bibnamefont{Aharony}},
  \emph{\bibinfo{title}{{Introduction to Percolation Theory}}}
  (\bibinfo{publisher}{Taylor and Francis, London}, \bibinfo{year}{1994}).

\bibitem[{\citenamefont{Melchert and Hartmann}(2008)}]{melchert2008}
\bibinfo{author}{\bibfnamefont{O.}~\bibnamefont{Melchert}} \bibnamefont{and}
  \bibinfo{author}{\bibfnamefont{A.~K.} \bibnamefont{Hartmann}},
  \bibinfo{journal}{New. J. Phys.} \textbf{\bibinfo{volume}{10}},
  \bibinfo{pages}{043039} (\bibinfo{year}{2008}).

\bibitem[{\citenamefont{Apolo et~al.}(2009)\citenamefont{Apolo, Melchert, and
  Hartmann}}]{apolo2009}
\bibinfo{author}{\bibfnamefont{L.}~\bibnamefont{Apolo}},
  \bibinfo{author}{\bibfnamefont{O.}~\bibnamefont{Melchert}}, \bibnamefont{and}
  \bibinfo{author}{\bibfnamefont{A.~K.} \bibnamefont{Hartmann}},
  \bibinfo{journal}{Phys. Rev. E} \textbf{\bibinfo{volume}{79}},
  \bibinfo{pages}{031103} (\bibinfo{year}{2009}).

\bibitem[{\citenamefont{Ahuja et~al.}(1993)\citenamefont{Ahuja, Magnanti, and
  Orlin}}]{ahuja1993}
\bibinfo{author}{\bibfnamefont{R.~K.} \bibnamefont{Ahuja}},
  \bibinfo{author}{\bibfnamefont{T.~L.} \bibnamefont{Magnanti}},
  \bibnamefont{and} \bibinfo{author}{\bibfnamefont{J.~B.} \bibnamefont{Orlin}},
  \emph{\bibinfo{title}{{Network Flows: Theory, Algorithms, and Applications}}}
  (\bibinfo{publisher}{Prentice Hall}, \bibinfo{year}{1993}).

\bibitem[{\citenamefont{Cook and Rohe}(1999)}]{cook1999}
\bibinfo{author}{\bibfnamefont{W.}~\bibnamefont{Cook}} \bibnamefont{and}
  \bibinfo{author}{\bibfnamefont{A.}~\bibnamefont{Rohe}},
  \bibinfo{journal}{INFORMS J. Computing} \textbf{\bibinfo{volume}{11}},
  \bibinfo{pages}{138} (\bibinfo{year}{1999}).

\bibitem[{\citenamefont{Hartmann and Rieger}(2001)}]{opt-phys2001}
\bibinfo{author}{\bibfnamefont{A.~K.} \bibnamefont{Hartmann}} \bibnamefont{and}
  \bibinfo{author}{\bibfnamefont{H.}~\bibnamefont{Rieger}},
  \emph{\bibinfo{title}{{Optimization Algorithms in Physics}}}
  (\bibinfo{publisher}{Wiley-VCH}, \bibinfo{address}{Weinheim},
  \bibinfo{year}{2001}).

\bibitem[{\citenamefont{Melchert}(2009{\natexlab{a}})}]{melchertThesis2009}
\bibinfo{author}{\bibfnamefont{O.}~\bibnamefont{Melchert}},
  \emph{\bibinfo{title}{{PhD thesis}}} (\bibinfo{publisher}{not published},
  \bibinfo{year}{2009}{\natexlab{a}}).

\bibitem[{\citenamefont{Hartmann}(2009)}]{practicalGuide2009}
\bibinfo{author}{\bibfnamefont{A.~K.} \bibnamefont{Hartmann}},
  \emph{\bibinfo{title}{{Practical Guide to Computer Simulations}}}
  (\bibinfo{publisher}{World Scientific}, \bibinfo{address}{Singapore},
  \bibinfo{year}{2009}).

\bibitem[{com()}]{comment_cookrohe}
\bibinfo{note}{For the calculation of minimum-weighted perfect matchings we use
  Cook and Rohes blossom4 extension to the Concorde library.},
  \urlprefix\url{http://www2.isye.gatech.edu/~wcook/blossom4/}.

\bibitem[{\citenamefont{Vachaspati and Vilenkin}(1984)}]{vachaspati1984}
\bibinfo{author}{\bibfnamefont{T.}~\bibnamefont{Vachaspati}} \bibnamefont{and}
  \bibinfo{author}{\bibfnamefont{A.}~\bibnamefont{Vilenkin}},
  \bibinfo{journal}{Phys. Rev. D} \textbf{\bibinfo{volume}{30}},
  \bibinfo{pages}{2036} (\bibinfo{year}{1984}).

\bibitem[{\citenamefont{Houdayer and Hartmann}(2004)}]{houdayer2004}
\bibinfo{author}{\bibfnamefont{J.}~\bibnamefont{Houdayer}} \bibnamefont{and}
  \bibinfo{author}{\bibfnamefont{A.~K.} \bibnamefont{Hartmann}},
  \bibinfo{journal}{Phys. Rev. {\bf B}} \textbf{\bibinfo{volume}{70}},
  \bibinfo{pages}{014418} (\bibinfo{year}{2004}).

\bibitem[{\citenamefont{Melchert}(2009{\natexlab{b}})}]{autoScale2009}
\bibinfo{author}{\bibfnamefont{O.}~\bibnamefont{Melchert}},
  \bibinfo{journal}{Preprint: arXiv:0910.5403v1}
  (\bibinfo{year}{2009}{\natexlab{b}}).

\bibitem[{\citenamefont{da~Silva et~al.}(2002)\citenamefont{da~Silva, Lyra, and
  Viswanathan}}]{dasilva2002}
\bibinfo{author}{\bibfnamefont{C.~R.} \bibnamefont{da~Silva}},
  \bibinfo{author}{\bibfnamefont{M.~L.} \bibnamefont{Lyra}}, \bibnamefont{and}
  \bibinfo{author}{\bibfnamefont{G.~M.} \bibnamefont{Viswanathan}},
  \bibinfo{journal}{Phys. Rev. E} \textbf{\bibinfo{volume}{66}},
  \bibinfo{pages}{056107} (\bibinfo{year}{2002}).

\bibitem[{\citenamefont{Jan et~al.}(1998)\citenamefont{Jan, Stauffer, and
  Aharony}}]{jan1998}
\bibinfo{author}{\bibfnamefont{N.}~\bibnamefont{Jan}},
  \bibinfo{author}{\bibfnamefont{D.}~\bibnamefont{Stauffer}}, \bibnamefont{and}
  \bibinfo{author}{\bibfnamefont{A.}~\bibnamefont{Aharony}},
  \bibinfo{journal}{J. Stat. Phys.} \textbf{\bibinfo{volume}{92}},
  \bibinfo{pages}{325} (\bibinfo{year}{1998}).

\bibitem[{\citenamefont{Melchert and Hartmann}(2009)}]{melchert2009}
\bibinfo{author}{\bibfnamefont{O.}~\bibnamefont{Melchert}} \bibnamefont{and}
  \bibinfo{author}{\bibfnamefont{A.~K.} \bibnamefont{Hartmann}},
  \bibinfo{journal}{Phys. Rev. B} \textbf{\bibinfo{volume}{79}},
  \bibinfo{pages}{184402} (\bibinfo{year}{2009}).

\bibitem[{\citenamefont{Aharony and Stauffer}(1995)}]{aharony1995}
\bibinfo{author}{\bibfnamefont{A.}~\bibnamefont{Aharony}} \bibnamefont{and}
  \bibinfo{author}{\bibfnamefont{D.}~\bibnamefont{Stauffer}},
  \bibinfo{journal}{Physica A} \textbf{\bibinfo{volume}{215}},
  \bibinfo{pages}{242} (\bibinfo{year}{1995}).

\bibitem[{\citenamefont{Schwartz et~al.}(1998)\citenamefont{Schwartz, Nazaryev,
  and Havlin}}]{schwartz1998}
\bibinfo{author}{\bibfnamefont{N.}~\bibnamefont{Schwartz}},
  \bibinfo{author}{\bibfnamefont{A.~L.} \bibnamefont{Nazaryev}},
  \bibnamefont{and} \bibinfo{author}{\bibfnamefont{S.}~\bibnamefont{Havlin}},
  \bibinfo{journal}{Phys. Rev. E} \textbf{\bibinfo{volume}{58}},
  \bibinfo{pages}{7642} (\bibinfo{year}{1998}).

\bibitem[{\citenamefont{Dobrin and Duxbury}(2001)}]{dobrin2001}
\bibinfo{author}{\bibfnamefont{R.}~\bibnamefont{Dobrin}} \bibnamefont{and}
  \bibinfo{author}{\bibfnamefont{P.~M.} \bibnamefont{Duxbury}},
  \bibinfo{journal}{Phys. Rev. Lett.} \textbf{\bibinfo{volume}{86}},
  \bibinfo{pages}{5076} (\bibinfo{year}{2001}).

\bibitem[{\citenamefont{Barab\'asi}(1996)}]{barabasi1996}
\bibinfo{author}{\bibfnamefont{A.~L.} \bibnamefont{Barab\'asi}},
  \bibinfo{journal}{Phys. Rev. Lett.} \textbf{\bibinfo{volume}{76}},
  \bibinfo{pages}{3750} (\bibinfo{year}{1996}).

\bibitem[{\citenamefont{Buldyrev et~al.}(2004)\citenamefont{Buldyrev, Havlin,
  L\'opez, and Stanley}}]{buldyrev2004}
\bibinfo{author}{\bibfnamefont{S.~V.} \bibnamefont{Buldyrev}},
  \bibinfo{author}{\bibfnamefont{S.}~\bibnamefont{Havlin}},
  \bibinfo{author}{\bibfnamefont{E.}~\bibnamefont{L\'opez}}, \bibnamefont{and}
  \bibinfo{author}{\bibfnamefont{H.~E.} \bibnamefont{Stanley}},
  \bibinfo{journal}{Phys. Rev. E} \textbf{\bibinfo{volume}{70}},
  \bibinfo{pages}{035102(R)} (\bibinfo{year}{2004}).

\bibitem[{\citenamefont{Middleton}(2000)}]{middleton2000}
\bibinfo{author}{\bibfnamefont{A.~A.} \bibnamefont{Middleton}},
  \bibinfo{journal}{Phys. Rev. B} \textbf{\bibinfo{volume}{61}},
  \bibinfo{pages}{14787} (\bibinfo{year}{2000}).

\bibitem[{\citenamefont{Cieplak et~al.}(1996)\citenamefont{Cieplak, Maritan,
  and Banavar}}]{cieplak1996}
\bibinfo{author}{\bibfnamefont{M.}~\bibnamefont{Cieplak}},
  \bibinfo{author}{\bibfnamefont{A.}~\bibnamefont{Maritan}}, \bibnamefont{and}
  \bibinfo{author}{\bibfnamefont{J.~R.} \bibnamefont{Banavar}},
  \bibinfo{journal}{Phys. Rev. Lett.} \textbf{\bibinfo{volume}{76}},
  \bibinfo{pages}{3754} (\bibinfo{year}{1996}).

\end{thebibliography}

\end{document}